\UseRawInputEncoding
\documentclass[twocolumn,superscriptaddress,aps, pra, reprint]{revtex4-2}

\usepackage{graphicx,amsfonts,amssymb,amsmath,hyperref,enumerate,footnote}
\usepackage{amsmath}
\usepackage{graphicx}
\usepackage{float}
\usepackage{indentfirst}
\usepackage{tikz}
\usetikzlibrary{quantikz}

\usepackage{gensymb}
\usepackage[utf8]{inputenc}
\usepackage{xcolor}
\usepackage{comment}

\newif\ifhyper
\hypertrue
\ifhyper
\hypersetup{
   citecolor = {green},
   colorlinks = {true}, 
   urlcolor = {blue} 
}
\fi

\newcommand{\beq}{\begin{equation}}
\newcommand{\eeq}{\end{equation}}
\newcommand{\beqa}{\begin{eqnarray}}
\newcommand{\eeqa}{\end{eqnarray}}

\def\ket#1{\vert#1\rangle}

\def\Longarrow{\protect\@lra}
\def\@lra{\relbar\joinrel\relbar\joinrel\relbar\joinrel%
          \relbar\joinrel\rightarrow}

\begin{document}

\title{Quantum-inspired clustering with light}

\author{Miguel Varga}
\affiliation{Centro de F\'isica de Materiales, UPV-EHU/CSIC, Paseo Manuel de Lardizabal 5, E-20018 San Sebastián, Spain}

\author{Pablo Bermejo}
\affiliation{Donostia International Physics Center, Paseo Manuel de Lardizabal 4, E-20018 San Sebasti\'an, Spain}

\author{Ruben Pellicer-Guridi}
\affiliation{Centro de F\'isica de Materiales, UPV-EHU/CSIC, Paseo Manuel de Lardizabal 5, E-20018 San Sebastián, Spain}
\affiliation{Donostia International Physics Center, Paseo Manuel de Lardizabal 4, E-20018 San Sebasti\'an, Spain}

\author{Rom\'an Or\'us}
\affiliation{Donostia International Physics Center, Paseo Manuel de Lardizabal 4, E-20018    San Sebasti\'an, Spain}
\affiliation{Ikerbasque Foundation for Science, Maria Diaz de Haro 3, E-48013 Bilbao, Spain}
\affiliation{Multiverse Computing, Paseo de Miram\'on 170, E-20014 San Sebasti\'an, Spain}

\author{Gabriel Molina-Terriza}
\affiliation{Centro de F\'isica de Materiales, UPV-EHU/CSIC, Paseo Manuel de Lardizabal 5, E-20018 San Sebastián, Spain}
\affiliation{Donostia International Physics Center, Paseo Manuel de Lardizabal 4, E-20018    San Sebasti\'an, Spain}
\affiliation{Ikerbasque Foundation for Science, Maria Diaz de Haro 3, E-48013 Bilbao, Spain}

\date{\today}

\begin{abstract}
This article introduces a novel approach to perform the simulation of a single qubit quantum algorithm using laser beams. Leveraging the polarization states of photonic qubits, and inspired by variational quantum eigensolvers, we develop a variational quantum  algorithm implementing a clustering procedure following the approach proposed by some of us in SciRep 13, 13284 (2023). A key aspect of our research involves the utilization of non-orthogonal states within the photonic domain, harnessing the potential of polarization schemes to reproduce unitary circuits. By mapping these non-orthogonal states into polarization states, we achieve an efficient and versatile quantum information processing unit which serves as a clustering device for a diverse set of datasets.
\end{abstract}

\maketitle

\emph{Introduction.-} In recent years, the field of quantum computing has witnessed a surge in novel proposals for quantum algorithms, many of which are strategically tailored to thrive in the Noisy Intermediate-Scale Quantum (NISQ) era, a period characterized by the presence of error-prone quantum hardware ~\cite{RevModPhys.94.015004,Preskill}. While quantum technology continues to evolve, the quest for more robust and reliable quantum algorithms remains paramount. Amidst this dynamic landscape, the ability to simulate quantum algorithms using classical computers for practical applications has not lost its relevance. As quantum algorithms diversify across different quantum computing architectures, a parallel trend is emerging: the development of increasingly efficient classical simulation methods. A good example of these methods are Tensor Networks \cite{TN}, which have proven recently capable of simulating the complex dynamics of many-qubit systems ~\cite{PRXQuantum.5.010308,PhysRevResearch.6.013326}. In addition, classical platforms are continually being innovated to replicate qubit behaviors, adding to the repertoire of simulation tools ~\cite{perez2015quantum, sun2022universal}. 

In this paper, we introduce an alternative approach to simulate variational quantum algorithms \cite{cerezo2021variational} by leveraging the capabilities of a photonic device as a dedicated processing unit. By using light, we can  replicate single-qubit rotations, therefore being able to implement tasks such as Variational Quantum Eigensolvers (VQE) \cite{VQE} and Variational Quantum Clustering (VQC) \cite{VQC}. Photonic quantum devices offer some inherent characteristics which make them compatible for such purpose: (i)  they present a processing unit which can be entirely mapped to the logical gates of a single-qubit circuit, (ii) they provide accurate control over noise, and (iii) they can be eventually scaled-up introducing actual quantum regimes for the input state. Using such architecture, we implement a photonic version of the quantum clustering algorithm in Ref.\cite{VQC} for a single qubit. This particular algorithm has  a simple structure that allows it to be implemented on NISQ devices, so that it can be run on a single qubit-like device as in this case. Therefore, our photonic classical device mimicks the behavior of a quantum circuit for a single qubit. Our results are a first test of the capabilities of such a classical platform to simulate quantum algorithms.

\bigskip 

\emph{Method.-} 
The algorithm at the core of this experiment is an unsupervised quantum clustering algorithm, designed for the classification of data points within a given dataset where no prior information about the system is available \cite{berry2019supervised}. In such a scenario there is no training stage, and the implementation follows that of Ref.\cite{VQC}, where a variational quantum circuit is optimized self-consistently so as to minimize the distance between points and cluster centroids. As explained in that reference, the procedure iteratively optimizes a set of variational parameters based on a reference cost function to determine the optimal configuration for the dataset. Notably, this approach relies solely on the intrinsic features of the dataset without any prior labeling.

To achieve this, the first step is to design a cost function. This cost function is built in such a way that its minimization provides the configuration of the optimal classification. The reference Hamiltonian, specifically constructed for this purpose, is given by:
\begin{equation}
    \mathcal{H} = \frac{1}{2}\sum_{i,j=1}^N  \left(d(\vec{x}_i,\vec{x}_j)+\lambda d(\vec{x}_i,\vec{c}_i)\right) \sum_{a = 1}^k \left(1-f_i^a\right)\left(1-f_j^a\right). 
    \label{eq1}
\end{equation}
In this equation, $N$ is the number of datapoints, $k$ is the number of clusters, subscripts $i,j$ represent each data point within the dataset, and $a$ denotes the label associated with each available cluster. The cost function takes into account the distance between data points $d(\vec{x}_i, \vec{x}_j)$, given by the l$^2$-norm, the distance between a data point and the centroid of its corresponding cluster $d(\vec{x}_i, \vec{c}_i)$, as well as the fidelity $f_i^a \equiv | \braket{\psi_i}{\psi^a} |^2$ between a variational quantum state $\ket{\psi_i}$ for datapoint $\vec{x}_i$ and a reference state $\ket{\psi^a}$ for cluster $a$. This fidelity is influenced by the position of each data point in the Poincaré polarization sphere, which is determined by a set of variational parameters. In addition, $\lambda$ serves as a regularization parameter, allowing for different penalizations of distances between dataset points and considering the relative importance of distances between data points and their respective cluster centroids. Indeed, the variational nature of the algorithm entirely falls within the function $f_i^a$. 

Our photonic implementation simulates the optimization of the above cost function for a single qubit, using a variational quantum circuit .  Our method consists of 2 distinct working units: a classical one, and a quantum-inspired one, which we simulate in our case with a diode laser. The classical computer will take care of upgrading the variational parameters driving the quantum circuit by means of a classical optimizer aiming at minimizing the cost function, as in a regular optimization problem. The quantum-inspired circuit will be then modulated based on the upgrade of the variational parameters, which will enter the circuit as rotations in the polarization of the light.

\bigskip

\emph{Experimental setup.-} To implement our quantum-inspired circuit, we have set a series of waveplates that modulate the polarization of an 808nm diode laser. These waveplates will adjust the laser beam's polarization based on variational parameters. Starting from a specific initial polarization state, the combination of waveplates will gradually transform it. The ultimate goal is to measure the polarization of a specific quantum state using a polarimeter, effectively allowing us to position polarized states throughout the Poincaré sphere, akin to the Bloch sphere of a single qubit. In essence, the system will serve as a versatile tool for transforming the positions of dataset points. It begins with an initial configuration and is manipulated to achieve an optimal configuration in which all points are situated on the surface of the Poincaré sphere. This minimizes a particular cost function, like the one in Eq.(\ref{eq1}). The underlying concept of using an 808nm diode laser stems from its simplicity and its capacity to map a single-qubit quantum circuit to the laser's operational principles. By replicating quantum logical gates through a combination of waveplates and facilitating the measurement process with a polarimeter, we can recreate the dynamics of the qubit. In our case, we use this setting to implement the photonic simulation of variational quantum clustering.  

\begin{figure}
    \centering
    \includegraphics[width=0.5\textwidth]{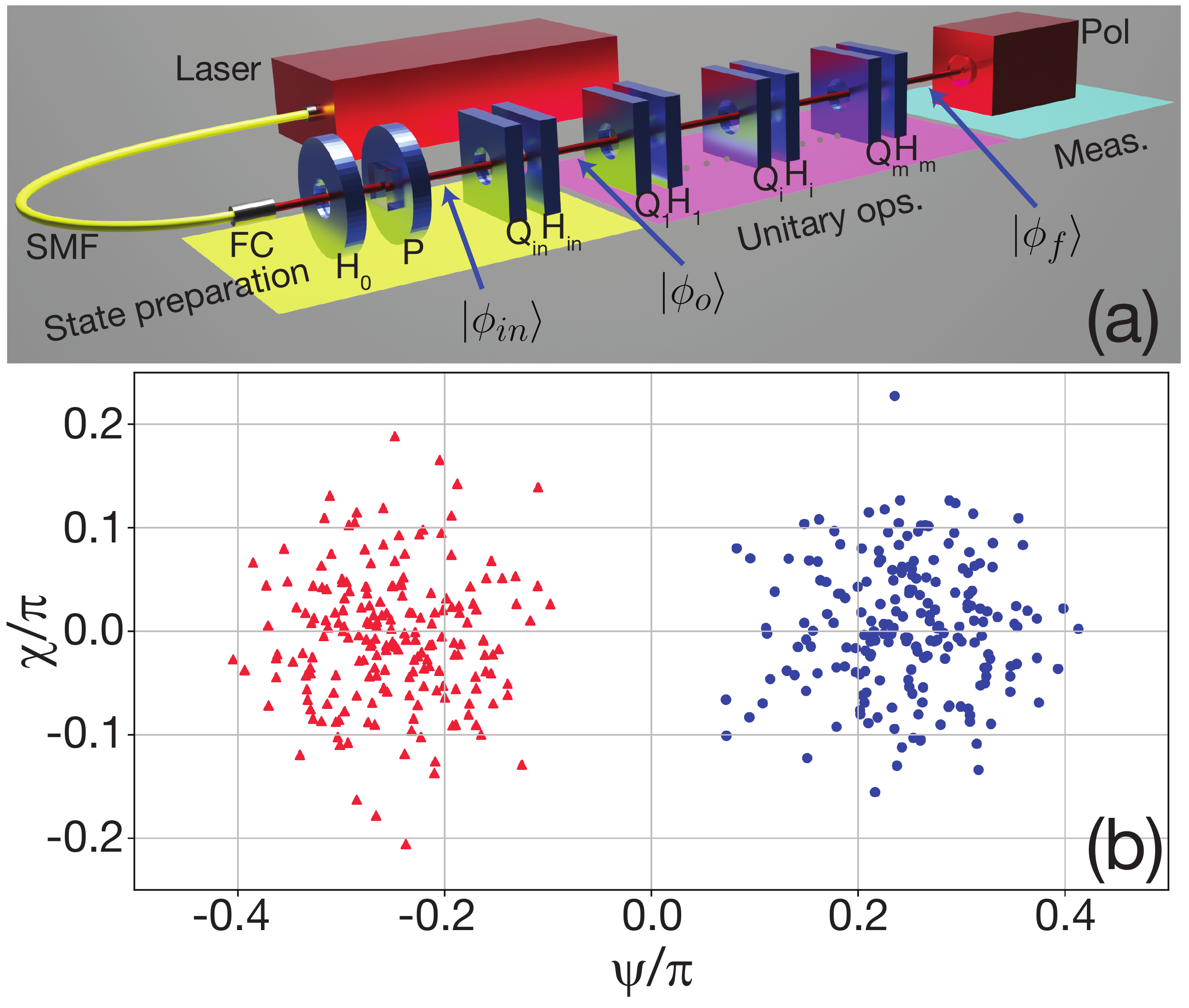}
    \caption{[Color figure] \emph{(a)} Scheme of optical setup. SMF: single mode optical fiber. FC: fiber collimator. P: linear polarizator. $\text{H}_\text{x}$: half wave plate. $\text{Q}_\text{x}$: quarter wave plate. Pol.: polarimeter. \emph{(b)} Example of cluster classification for a configuration of $2$ clusters of $200$ points defined by $\frac{d}{\sigma}=8$, where $d$ is the distance between centers and $\sigma$ the width of the Gaussian blobs.}
    \label{fig:setup}
\end{figure}

The optical setup used to initialize the states is shown in Fig.\ref{fig:setup}. In the state preparation sequence, after half wave plate $H_0$ and polarizer $P$ are applied in the incoming beam, the initial state of the system can be represented as 
\begin{equation}
    \ket{\phi_{in}} = \ket{h} \equiv \left(\begin{array}{r}
    1 \\
     0
  \end{array}\right),
\end{equation}
which corresponds to horizontal polarization. This configuration, $H_0 + P$, is fixed in order to allow maximum intensity and a stable polarization.

The two first plates, $Q_{in}$ and $H_{in}$, are used to transform the states from the data feature space to the Poincaré sphere. They are mounted on separate rotary stages (RSW60C-T3A from Zaber). These motors have a maximum accuracy of $0.08\degree$ and a maximum speed of $450\degree/s$. The state $|\phi_o\rangle$, defined by the angles $(\psi,\chi)$, is determined by the angles $(\alpha, \beta)$ corresponding to the fast axis orientation of $Q_{in}$ and $H_{in}$, respectively. This bijection was done by means of a look-up-table.

After initialization, the state $\ket{\phi_o}$ is modified by subsequent sets of $m$ half and quarter wave plates $\{\hat{H}_k,\hat{Q}_k\}$. These wave plates work in the same way as the initial ones, that is, for ideal waveplates their Jones matrices are given by 
\begin{equation}\label{eq:half}
\small
    \hat{H}_k = e^{-\frac{i\pi}{2}}\left (
  \begin{array}{cc}
    \cos^2(\beta_k) - \sin^2(\beta_k) & 2\cos(\beta_k)\sin(\beta_k) \\
     2\cos(\beta_k)\sin(\beta_k) & \sin^2(\beta_k) - \cos^2(\beta_k) 
  \end{array}
\right )
\end{equation}
for the half wave plates, and
\begin{equation}\label{eq:quarter}
\small
    \hat{Q}_k = e^{-\frac{i\pi}{4}}\left (
  \begin{array}{cc}
    \cos^2(\alpha_k) +i\sin^2(\alpha_k) & (1-i)\cos(\alpha_k)\sin(\alpha_k) \\
     (1-i)\cos(\alpha_k)\sin(\alpha_k) & \sin^2(\alpha_k) +i\cos^2(\alpha_k) 
  \end{array}
\right )
\end{equation}
for the quarter wave plates. While experimentally, the waveplates can depart from this idealized model, this allows us to simulate the behaviour of our experimental system. Here, $\beta_k$ and $\alpha_k$ are the angle of rotation of the waveplates. They act as the variational parameters to be optimized in the variational circuit optimizing the cost function from Eq.(\ref{eq1}). Therefore, the final state $\ket{\phi_f}$ of the system is given by
\begin{equation}
    \ket{\phi_f} = \hat{H}_m\hat{Q}_m\dots\hat{H}_2\hat{Q}_2\hat{H}_1\hat{Q}_1|\psi_o\rangle,
\end{equation}
or, in terms of the initial horizontal polarization state $\ket{\psi_{in}}$, 
\begin{equation}
    \ket{\phi_f} = \hat{H}_m\hat{Q}_m\dots\hat{H}_2\hat{Q}_2\hat{H}_1\hat{Q}_1\hat{H}_{in}\hat{Q}_{in}\ket{\psi_{in}}.
\end{equation}
These output states $\ket{\phi_o}$ are directly measured by the polarimeter. In the case of upgrading this experiment to more qubits, the polarimeter should be replaced by a complete tomography of the output state. In our case, this is simplified as the polarimeter provides the value of the Stokes parameters $s_0, s_1, s_2, s_3$ corresponding to a point in the Poincaré polarization sphere.  The readout of this point in the sphere allows for the self-consistent optimization of the variational parameters in the circuit, allowing in turn to minimize the clustering cost function and therefore implement an unsupervised classification of the points in the dataset. 

The results for the case of two clusters with  $\sim 100$ points are shown in Fig. \ref{fig:setup}(b). One can observe that the algorithm successfully automatically classifies the points in the two different clusters. The ratio of success of the algorithm for this particular case is of $100\%$. In this case, the classification task is conducted on top of a random exploration of the initialization step, in order to favour the exploration of the parameters space. 

\bigskip

\begin{figure}[t]
  \centering
  \includegraphics[width=0.5\textwidth]{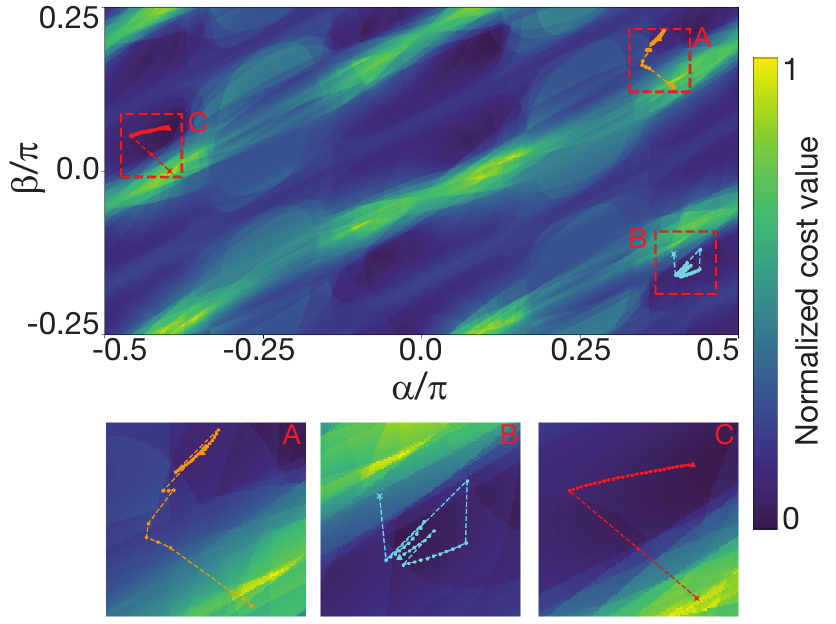}
  \caption{[Color online] Numerical example for the clusterization of a 4 clusters distribution. \emph{Top:} Colormap of the cost function value with respect of the hiperparameters $\alpha$ and $\beta$ corresponding to the rotation angle of a half and a quarter wave-plate, respectively. \emph{Bottom:} insets' detail of the trajectories taken to get to the local minima of the cost function.}
  \label{fig:trajectories}
\end{figure}

\emph{Numerical analysis.-} To better understand the phase space of the variational parameters and provide a better intuition for more complex quantum optimization processes, we performed a numerical analysis for a $4$ Gaussian blobs dataset, classified with $2$ single variational layers, generating 3 different optimization paths in hyperparameter space (as shown in Fig.(\ref{fig:trajectories}a,\ref{fig:trajectories}b, \ref{fig:trajectories}c). We present in the figure the landscape of the cost function, which as expected is a complex shape with many local minima. The initialization of the algorithm would start at a random point in the landscape and then, subsequently, the optimization algorithm would provide the rotations of the Poincare sphere which would optimize the cost function, providing a path in the landscape.

\begin{figure}[t]
  \centering
  \includegraphics[width=0.5\textwidth]{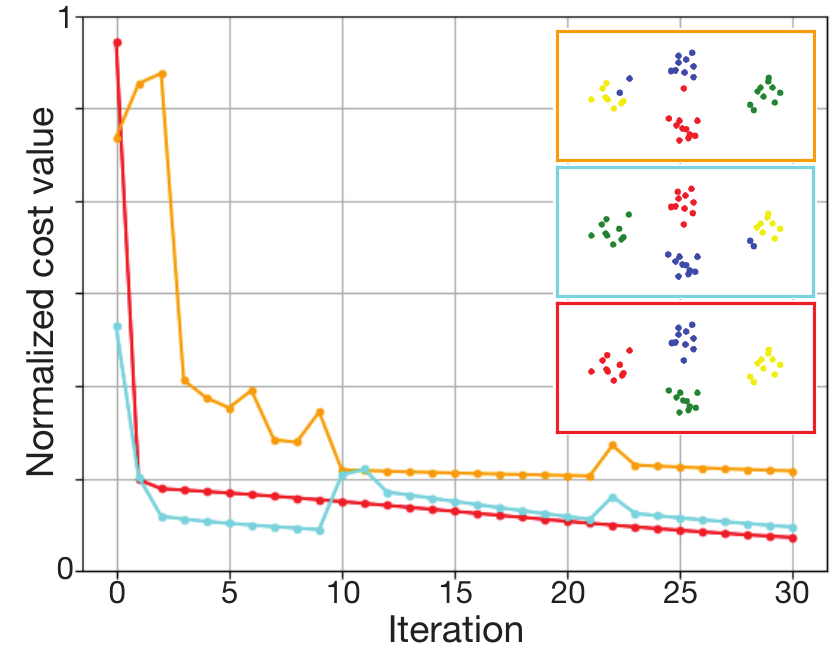}
  \caption{Numerical cost function evolution for 4 Gaussian blobs: 3 different trajectories corresponding to the insets of the figure \ref{fig:trajectories}}.
  \label{fig:cost}
\end{figure}

Notice that while there is a single absolute minima, most of the local minima also provide a good classification. The three examples shown fall into the local minima with success ratios of $92.5\%$, $95\%$ and $100\%$. The final classification can be shown in Fig.\ref{fig:cost} with the corresponding evolution of the cost function. One can observe that, after just $10$ iterations, the  cost function typically arrives to a stable solution, while it may need up to $30$ iterations to reach the minimum. Only one of the optimization paths ends up displaying perfect classification, corresponding to the one with the smallest cost. This result is a consequence of how sensible variational algorithms are to initialization ~\cite{truger2023warm, egger2021warm}, which has become one of the main features to look at in the search for non-classical simulability of variational algorithms \cite{cerezo2023does}. 

As mentioned earlier, one could use more complex minimization algorithms, but in our case a combination of Monte Carlo and steepest descent has provided good results. In this work, we focus on the capabilities of the method to reproduce efficiently a single qubit algorithm, showing the main features of this clustering scheme and opening the way for further tuning strategies.

\bigskip

\begin{figure}
    \centering
    \includegraphics[width=0.5\textwidth]{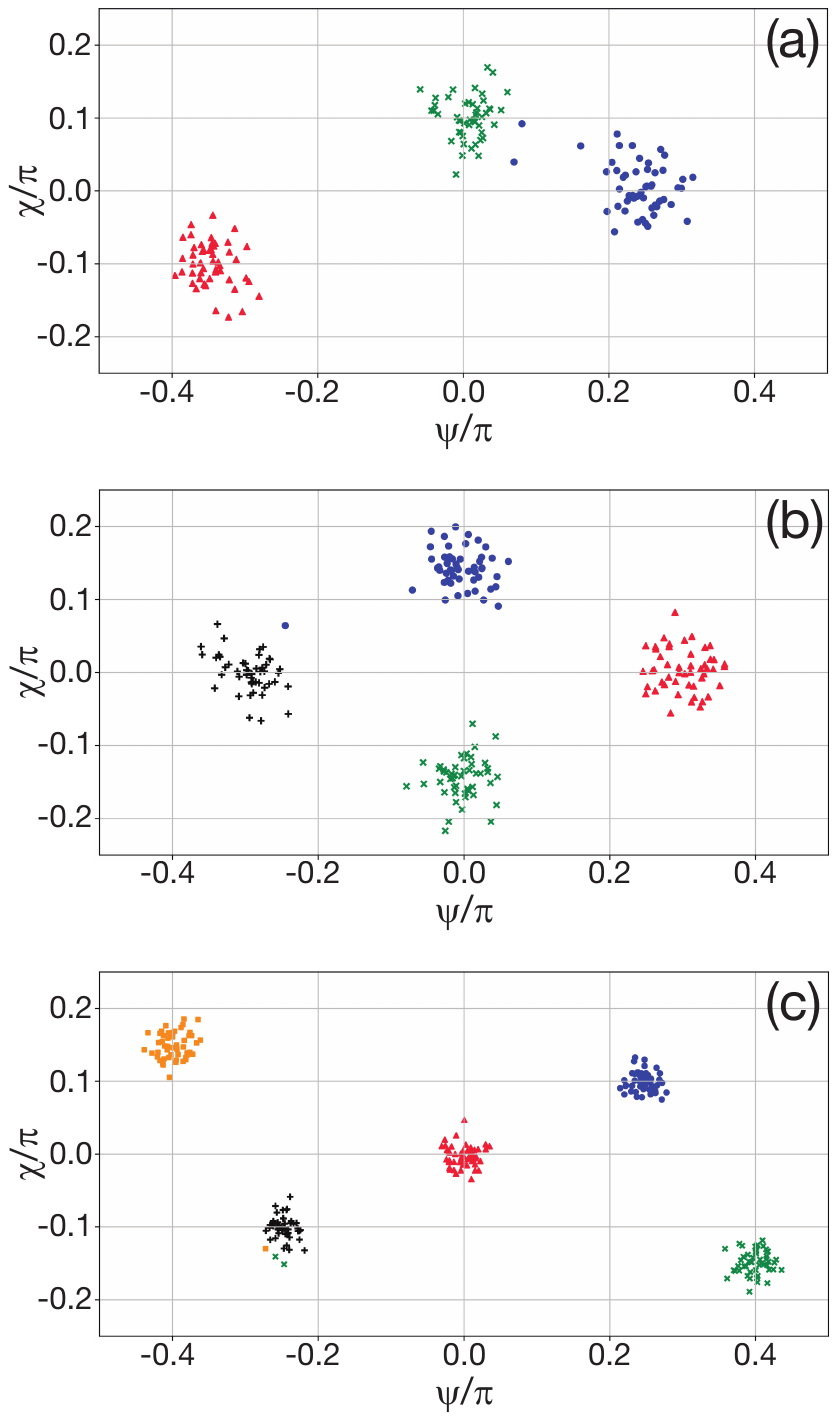}
    \caption{[Color online] Experimental clustering results: (a) 3 gaussian blobs, (b) 4 gaussian blobs and (5) 5 gaussian globes. Colors correspond to the different clusters identified by the experiment.}
    \label{fig:results_a}
\end{figure}

\begin{figure}
    \centering
    \includegraphics[width=0.5\textwidth]{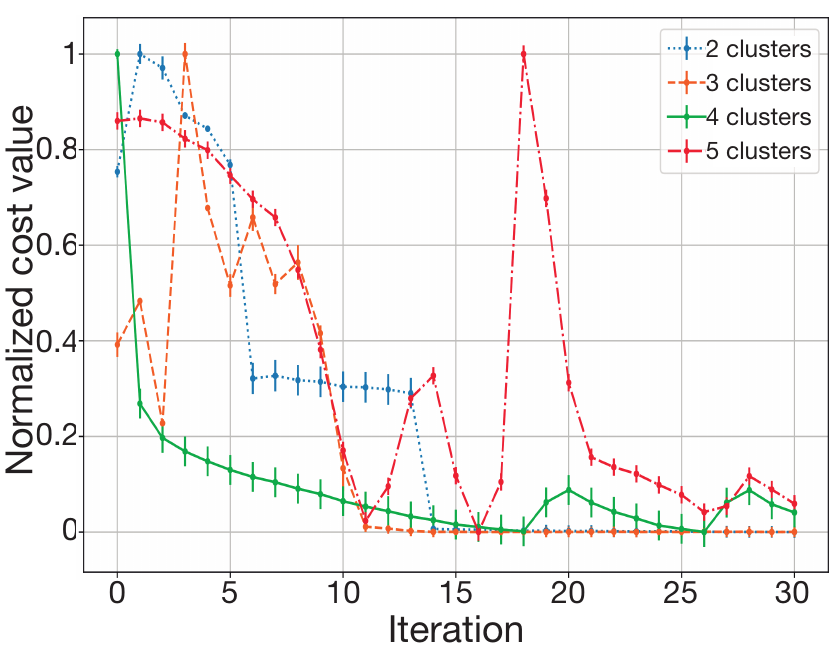}
    \caption{[Color online] Experimental cost function evolution corresponding to the distributions of the figures \ref{fig:setup} and \ref{fig:results_a}. The cost values where normalized between $0$ and $1$.}
    \label{fig:cost_evolution_exp}
\end{figure}

\emph{Experimental results.-} The methods described before were implemented in our clustering experiment, using more clusters, in order to test the experimental limitations of the system. The main results are summarized in the plots in Fig.\ref{fig:results_a}. Similarly as in Fig. \ref{fig:setup}(b) the figure shows the capability of the photonic clustering implemented to automatically identify configurations with different quantity of clusters. Additionally, in figure \ref{fig:cost_evolution_exp} we provide the experimental evolution of the cost function for the for different configurations tested. These results can be compared with the numerical results that we provided earlier. It can be observed that the experimental errors do not significantly affect the expected behaviour of the optimization process. 

The results presented above constitute a first proof of principle of the validity of using classical photonic systems to simulate quantum clustering. The experiment can be further expanded in many different directions. For instance, it should be possible to classify more complex datasets. In addition, one may also explore the possibility of simulating a multi-qubit quantum circuit, by generating and manipulating quantum states consisting of more than one photon.

\bigskip 

\emph{Conclusions and outlook.-} In this paper we have introduced a novel quantum-inspired clustering method, based on the photonic simulation of single-qubit dynamics. We have shown how it is possible to optimize the polarization degree of freedom of a laser beam, in the same way that one can optimize the quantum state of a single qubit, so as to minimize a clustering cost function. We have implemented the experiment and shown the validity of our ideas, performing automatic clustering of random data, with no prior information, scattered in up to five zones, with perfect accuracy. 

Our experiment is based on encoding the information on the polarization of light, akin to using a single qubit in a quantum algorithm. The rotation matrix introduced by the wave platesin our experiment, can be cast in the quantum circuit in terms of general Pauli rotations around the main axis of a qubit. This can be useful in order to build a bridge between this specific implementation and the virtual environments commonly used in academia and industry with access to quantum hardware (Pennylane, Qiskit, etc...) ~\cite{bergholm2018pennylane, Qiskit}. There exist several restrictions in the amount and type of logical gates allowed in actual quantum hardware, so a mapping between the functioning of this optical circuit and universal sets of gates is advisable. 

Our work can be further expanded in many directions, as discussed previously. We believe that a promising path is the simulation of multi-qubit systems. In addition, the flexibility of the variational circuit allows for a wide variety of applications. With this in mind, one could for instance build up diverse cost functions for different purposes using the same experimental arrangement, so that we could indeed use a laser beam to implement different types of quantum-inspired machine learning strategies. Indeed, when writing up this paper, we noticed the proposal of a very similar similar set-up for the realization of VQE algorithms using photonic devices \cite{stornati2023variational}. Last but not least, our scheme allows the introduction of other features, such as data reuploading ~\cite{reuploading, schuld2021effect}. Data reuploading was already proposed in similar contexts, such as in Ref.\cite{jerbi2023quantum}, where this was used to outperform kernel methods using very few quantum resources, as low as one single qubit. Even without reuploading, introducing several layers of gates in the circuit helps in practice in the convergence of the variational optimization, allowing for small changes of the different parameters at each iteration.  All these topics will be the subject of future investigations. 

{\bf Acknowldgements.-} We acknowledge Donostia International Physics Center (DIPC), Centro de F\'isica de Materiales, Ikerbasque, Basque Government, Diputaci\'on de Gipuzkoa, Spanish Ministry of Science and Innovation and European Innovation Council (EIC)  for constant support, as well as insightful discussions with the teams from Multiverse Computing, DIPC and CFM on the algorithms and technical implementations. This work was supported by the Spanish Ministry of Science and Innovation through the PLEC2021-008251  project. We also acknowledge support from PTI-001 from CSIC and the Ministry of Science through project PID2022-143268NB-I00.

\bibliographystyle{apsrev4-2}
\bibliography{biblio}

\end{document}